\documentclass[8pt,preprint]{aastex}

\usepackage{epsf}
\usepackage{amssymb}
\usepackage{graphicx}
\usepackage{lscape}

\newcommand{\kmps}{km s$^{-1}$} 
\newcommand{\Rsun}{R$_{\odot}$} 
\newcommand{\Msun}{M$_{\odot}$} 
\newcommand{\dechms}[4]{$#1^{\rm h}#2^{\rm m}#3\mbox{$^{\rm s}\mskip-7.6mu.\,$}#4$} 
\newcommand{\decdms}[4]{$#1^{\circ}#2'#3\mbox{$''\mskip-7.6mu.\,$}#4$} 
\newcommand{\mdeg}[2]{$#1\mbox{$^\circ \mskip-7.6mu.\,$}#2$} 
\newcommand{\msec}[2]{$#1\mbox{$''\mskip-7.6mu.\,$}#2$}
\newcommand{\mmsec}[2]{$#1\mbox{$^s\mskip-7.6mu.\,$}#2$}

\begin{document}

\title{VLBA determination of the distance to nearby star-forming regions\\
  V. Dynamical mass, distance and radio structure of V773~Tau~A}

\author{Rosa M. Torres,}

\affil{Argelander-Institut f\"ur Astronomie, Universit\"at Bonn, Auf
  dem H\"ugel 71, 53121 Bonn, Germany. \email{rtorres@astro.uni-bonn.de}}

\author{Laurent Loinard\altaffilmark{1},}

\affil{Max-Planck-Institut f\"ur Radioastronomie, Auf dem H\"ugel 69, 53121 Bonn, Germany}

\author{Amy J. Mioduszewski,}

\affil{Dominici Science Operations Center, National Radio Astronomy
  Observatory, 1003 Lopezville Road, Socorro, NM 87801.}

\author{Andrew F. Boden,}

\affil{Division of Physics, Math, and Astronomy, California
  Institute of Technology, 1200 E California Blvd., Pasadena CA 91125.}

\author{Ramiro Franco-Hern\'andez\altaffilmark{2}, Wouter H. T. Vlemmings\altaffilmark{3},}

\affil{Argelander-Institut f\"ur Astronomie, Universit\"at Bonn, Auf
  dem H\"ugel 71, 53121 Bonn, Germany.}

\and

\author{ Luis F. Rodr\'{\i}guez\altaffilmark{4}}

\affil{Centro de Radiostronom\'{\i}a y Astrof\'{\i}sica, Universidad
  Nacional Aut\'onoma de M\'exico, Apartado Postal 72--3 (Xangari), 58089
  Morelia, Michoac\'an, M\'exico.}

\altaffiltext{1}{Centro de Radiostronom\'{\i}a y Astrof\'{\i}sica, Universidad
  Nacional Aut\'onoma de M\'exico, Apartado Postal 72--3 (Xangari), 58089
  Morelia, Michoac\'an, M\'exico.}

\altaffiltext{2}{Departamento de Astronom\'{\i}a, Universidad de Chile, Casilla 
36-D, Santiago, Chile}

\altaffiltext{3}{Onsala Space Observatory, SE-439 92 Onsala, Sweden}

\altaffiltext{4}{Astronomy Department, Faculty of Science, King Abdulaziz 
University, P.O. Box 80203, Jeddah 21589, Saudi Arabia}

\begin{abstract}
  We present multi-epoch Very Long Baseline Array (VLBA) observations of
  V773~Tau~A, the 51-day binary subsystem in the multiple young stellar system
  V773~Tau. Combined with previous interferometric and radial velocity
  measurements, these new data enable us to improve the characterization of
  the physical orbit of the A subsystem. In particular, we infer updated
  dynamical masses for the primary and the secondary components of 1.55 $\pm$
  0.11 \Msun, and 1.293 $\pm$ 0.068 \Msun, respectively, and an updated
  {\it orbital parallax} distance to the system of 135.7 $\pm$ 3.2 pc, all consistent 
  with previous estimates. Using the improved orbit, we can calculate the absolute
  coordinates of the barycenter of V773~Tau~A at each epoch of our VLBA
  observations, and fit for its {\it trigonometric parallax} and proper motion. This
  provides a direct measurement of the distance to the system almost entirely
  independent of the orbit modeling. The best fit yields a distance of 129.9
  $\pm$ 3.2 pc, in good agreement (i.e.\ within 1$\sigma$) with the distance
  estimate based on the orbital fit. Taking the mean value of the {\it orbital}
  and trigonometric parallaxes, we conclude that V773~Tau is located at $d$ =
  132.8 $\pm$ 2.3 pc. The accuracy of this determination is nearly one order of
  magnitude better than that of previous estimates.  In projection, V773~Tau and 
  two other young stars (Hubble~4 and HDE~283572) recently observed with the 
  VLBA are located toward the dark cloud Lynds 1495, in the central region of 
  Taurus. These three stars appear to have similar trigonometric parallaxes, radial
  velocities, and proper motions, and we argue that the weighted mean and 
  dispersion of their distances ($d$ = 131.4 pc and $\sigma_d$ = 2.4 pc) provide a 
  good estimate 
  of the distance to and depth of Lynds~1495 and its associated stellar population. 
  The radio emission from the two sources in V773~Tau~A is largely of gyrosynchrotron
  origin. Interestingly, both sources are observed to become typically five
  times brighter near periastron than near apastron (presumably because of
  increased flaring activity), and the separation between the radio sources
  near periastron appears to be systematically smaller than the separation
  between the stars. While this clearly indicates some interaction between the
  individual magnetospheres, the exact mechanisms at play are unclear because
  even at periastron, the separation between the stars ($\sim$ 30 $R_*$) remain
  much larger than the radius of the magnetospheres around these low-mass
  young stars ($\sim$ 6 $R_*$).
\end{abstract}


\section{Introduction}
In spite of significant progress in recent years, the formation and early
evolution of stars are still not fully understood (e.g.\ Hillenbrand \& White
2004, Mathieu et al.\ 2007). One way to foster progress is to measure as
accurately as possible the intrinsic characteristics of individual young stars
(luminosity, effective temperature, mass, disk properties, etc.), and compare
them with the predictions of detailed theoretical models. Young binary systems
are particularly interesting in this respect, because tracking their orbital
motions provides a direct means to estimate their dynamical mass. In
particular, if astrometric and spectroscopic data are combined, the physical
orbit and the individual masses of the system members can be determined.

A recurrent obstacle to the accurate determination of the intrinsic properties
of young stars have been fairly large uncertainties (typically 20 to 50\%) in
distance estimates to even the nearest star-forming regions. Significant
progress has been possible in recent years thanks to direct trigonometric
parallax measurements obtained using Very Long Baseline Interferometry (VLBI)
multi-epoch observations. Such results have been reported, in particular, in
the previous papers in this series (Loinard et al.\ 2005, 2007, 2008; Torres
et al.\ 2007, 2009; Dzib et al.\ 2010, 2011). They provide uncertainties of a few
percent or better, that typically surpass the accuracy of previous
determinations by one order of magnitude.

V773~Tau (HD~283447, HBC~367) is a young stellar system located toward the
dark cloud Lynds~1495 in Taurus. V773~Tau was first identified as a T~Tauri
star by Rydgren et al.\ (1976), and was established as a visual double
(components designated A and B) with an apparent separation of roughly 150 mas
in high-angular resolution studies independently by Ghez et al.\ (1993) and
Leinert et al.\ (1993). Roughly contemporaneously, Martin et al.\ (1994) and
Welty (1995) established the brighter (A) visual component as a short-period
(51 days) double-lined spectroscopic binary. Duch\^ene et al.\ (2003; hereafter
D2003) and Woitas (2003) independently identified a third visual component
(designated C; note that D2003 use an alternate component notation) in the
system, indicating that V773~Tau is (at least) a quadruple system.\footnote{It
  should be noted that multiple nomenclatures have been used to describe the
  components of V773~Tau. D2003 designate the spectroscopic binary
  V773~Tau~A/B, and the optical and infrared companions V773~Tau~C and
  V773~Tau~D, respectively. Alternately White \& Ghez (2001), Woitas (2003),
  and Massi et al.\ (2008) among others, designate the spectroscopic binary as
  V773~Tau~A (containing stars Aa and Ab), and the two companions as B and C,
  respectively. A third nomenclature is used in the Double Star Catalog of
  Mason et al.\ (2001), which designates the spectroscopic binary Aa,
  containing stars Aa1 and Aa2. In this work we will follow the notation
  established by White \& Ghez (2001), reflecting the hierarchical nature of
  the V773~Tau architecture.}

V773~Tau~A has long been known to be a strong radio source (Kutner et al.\
1986).\footnote{The two companions V773~Tau~B and C, on the other hand, are
  not detected at radio wavelengths at the level of sensitivity of existing
  observations.} Indeed, it was the strongest source in the 5 GHz VLA survey
of WTTS in the Taurus-Auriga molecular cloud complex by O'Neal et al.\
(1990). From detailed multi-frequency observations, Feigelson et al.\ (1994)
concluded that the radiation was most likely of non-thermal origin. This was
confirmed by Phillips et al.\ (1996; hereafter P1996) who obtained VLBI
observations, and resolved the radio emission into a clear double source, most
likely corresponding to the two components of the spectroscopic binary. More
recently, Massi et al.\ (2002, 2006) showed that the radio emission exhibits
periodic variations with a period corresponding to the 51 day orbital period
of the spectroscopic binary. This variability is due to an increase in the
flaring activity near periastron and might reflect interactions between
extended magnetic structures associated with the two stars when they get close
to one another. Finally, Boden et al.\ (2007; hereafter B2007) and Massi et
al.\ (2008) also resolved the radio emission from V773~Tau~A into two
components, which they associate with the two stars in the spectroscopic
binary.

The dynamics in V773~Tau has been studied by a number of authors. Relative
orbital motion among the A, B, C components has been
monitored by D2003, B2007 and Boden et al.\ (2011; a companion paper to this
one -- hereafter B2011). Following Welty (1995) and P1996, B2007 used radio
and near-IR interferometry and spectroscopic radial velocity (RV) datasets to
estimate the A subsystem physical (three-dimensional) orbit. They obtain
dynamical mass estimates of 1.54 and 1.33 \Msun\ for the Aa and Ab components,
respectively. Further, B2007 estimated the A subsystem distance by means of
``orbital parallax'' (comparing the angular and physical orbit size), yielding
136.2 $\pm$ 3.7 pc. There is also a direct trigonometric parallax measurement
based on multi-epoch VLBI observations for this source (Lestrade et al.\
1999). This VLBI-based distance measurement ($d$ = 148.4$^{+5.7}_{-5.3}$ pc)
is roughly consistent (at the 2--3 $\sigma$ level) with the value obtained
from the A subsystem orbit modeling.

In this paper we present new VLBI observations of V773~Tau~A which resolve the
A subsystem, and yield additional insights on its structure. These data are
described in \S 2, and modeled jointly with earlier observations from B2007 
in \S3 to update the A subsystem physical orbit and resulting physical parameters 
(component dynamical masses, orbital parallax). Then the updated A orbit model 
is used in conjunction with the VLBA global astrometry to compute a new trigonometric 
parallax to V773~Tau~A. The relevance of these new results for the distance to the 
dark cloud Lynds~1495 is discussed in \S~3.3. Several interesting features of the 
V773~Tau~A radio emission are apparent in these new observations; the insights 
that they yield on the magnetospheric physics of the A subsystem components 
are discussed in \S~4.2.

{\scriptsize 
\begin{deluxetable}{clcr}
\tablecaption{Project code and date for each observation.\label{table_observations}}
\tablehead{
\multicolumn{1}{c}{Epoch}&
\multicolumn{1}{c}{Project Code}&
\multicolumn{1}{c}{Mean UT date}&
\multicolumn{1}{c}{Julian Day}\\
{}&
{}&
{[yyyy.mm.dd ~ hh:mm]}&
{}}%
\startdata
01 & BM 198 A  & 2004.03.11 ~ 20:12 & 2453076.3417 \\
02 & BM 198 B  & 2004.03.12 ~ 20:12 & 2453077.3417 \\
03 & BM 198 C  & 2004.03.13 ~ 20:12 & 2453078.3417 \\
04 & BM 198 D  & 2004.03.14 ~ 20:12 & 2453079.3417 \\
05 & BM 198 E  & 2004.03.15 ~ 20:11 & 2453080.3414 \\
06 & BM 198 F  & 2004.03.16 ~ 20:11 & 2453081.3414 \\
07 & BM 198 G  & 2004.03.17 ~ 20:11 & 2453082.3414 \\
08 & BL 128 AA & 2005.09.08 ~ 12:01 & 2453622.0013 \\
09 & BL 128 AB & 2005.11.15 ~ 07:31 & 2453689.8136 \\
10 & BL 128 AC & 2006.01.21 ~ 03:11 & 2453756.6327 \\
11 & BL 128 AD & 2006.04.01 ~ 22:31 & 2453827.4386 \\
12 & BL 128 AE & 2006.06.12 ~ 17:48 & 2453899.2423 \\
13 & BL 128 AF & 2006.09.05 ~ 12:14 & 2453984.0102 \\
14 & BL 146 B  & 2007.08.23 ~ 13:06 & 2454336.0461 \\
15 & BL 146 C  & 2007.08.29 ~ 12:42 & 2454342.0295 \\
16 & BL 146 D  & 2007.09.05 ~ 12:15 & 2454349.0106 \\
17 & BL 146 E  & 2007.09.11 ~ 11:51 & 2454354.9943 \\
18 & BL 146 F  & 2007.09.16 ~ 11:32 & 2454359.9806 \\
19 & BL 146 G  & 2007.09.21 ~ 11:12 & 2454364.9669 \\
20 & BL 146 H  & 2007.09.27 ~ 10:48 & 2454370.9504 \\
21 & BL 146 I  & 2007.10.03 ~ 10:25 & 2454376.9342 \\
22 & BL 146 J  & 2007.10.09 ~ 10:01 & 2454382.9177 \\
23 & BL 146 K  & 2007.10.17 ~ 09:30 & 2454390.8960 \\
24 & BL 146 L  & 2007.10.23 ~ 09:06 & 2454396.8794 \\
25 & BL 146 M  & 2007.10.27 ~ 08:50 & 2454400.8684 \\
26 & BL 146 N  & 2007.11.17 ~ 07:28 & 2454421.8114 \\
27 & BM 306    & 2009.09.27 ~ 08:07 & 2455101.8386 \\
\enddata
\end{deluxetable}
}

\section{Observations and data calibration}

In this article, we will make use of a total of 27 continuum observations of
V773~Tau~A obtained with the VLBA at $\lambda$ = 3.6 cm (see Table
\ref{table_observations} for details). These observations correspond to four
different projects (labeled BM198, BL128A, BL146, and BM306) that we will now
describe separately.

\subsection{BM198}

This first series of observations is archival (P.I.: M.\ Massi), and
corresponds to the data published initially by B2007. It consists of seven
observations (epochs 1--7 in Table \ref{table_observations}) obtained every
day between March 11 and March 17, 2004, when the system was near
apastron. The phase center was at $\alpha_{J2000.0}$ =
\dechms{04}{14}{12}{9198}, $\delta_{J2000.0}$ = +\decdms{28}{12}{12}{199}. Six
hours of telescope time were allocated to each epoch, and each observation
consisted of series of cycles with four minutes spent on source, and two
minutes spent on the main phase-referencing quasar J0403+2600, located
\mdeg{3}{31} away. Every 30 minutes, the secondary calibrator J0408+3032 was
also observed. The data were collected with the VLBA plus Effelsberg, but for
consistency with the rest of the data used in this paper (which were all
VLBA only), the Effelsberg antenna was not included during this data reduction.

\begin{figure}[!t]
  \centerline{\includegraphics[height=0.45\textwidth,angle=0]{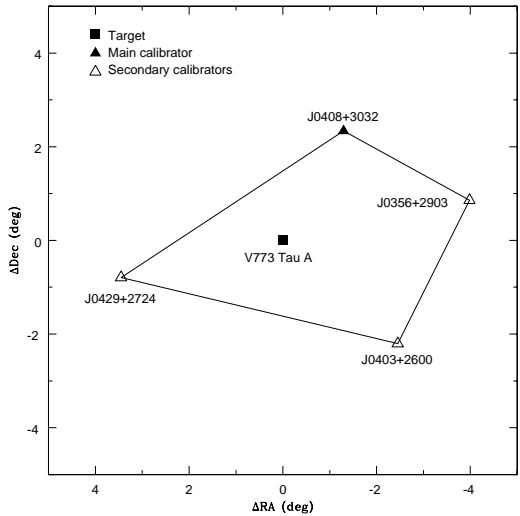}}
  \caption{Relative position of the astronomical target, the main calibrator
    (J0408+3032), and the secondary calibrators (J0403+2600, J0429+2724, and
    J0356+2903) for projects BL128A, BL146, and
    BM306.\label{fig_configuration}}
\end{figure}

\subsection{BL128A}

The second set of data consists of 6 observations obtained every two to three
months between September 2005 and September 2006 (epochs 8--13 in Table
\ref{table_observations}), and was initially designed to follow the
parallactic path of the source. The phase center was at $\alpha_{J2000.0}$ =
\dechms{04}{14}{12}{9195}, $\delta_{J2000.0}$ = +\decdms{28}{12}{12}{291}. Two
hours of telescope time were allocated to each epoch. Each observation
consisted of series of cycles with two minutes spent on source, and one minute
spent on the main phase-referencing quasar J0408+3032, located \mdeg{2}{67}
away. Every 24 minutes, we also observed three secondary calibrators
(J0403+2600, J0429+2724, and J0356+2903) which, together with J0408+3032, form
a lozenge around the astronomical source (Figure \ref{fig_configuration}). All
four calibrators are very compact extragalactic sources whose absolute
positions are known to better than 0.58, 0.40, 0.41, and 1.46 mas,
respectively (taken from L. Petrov, solution 2008, unpublished, available at
http://vlbi.gsfc.nasa.gov/).

\subsection{BL146}

This third data set corresponds to 13 observations obtained roughly every 7
days between August 29 and November 17, 2007 (epochs 14--26 in Table
\ref{table_observations}). These observations span a total of 82 days (about
1.6 orbital periods of the system), and were designed to follow the orbital
motion over a complete revolution. In this case, the phase center was at
$\alpha_{J2000.0}$ = \dechms{04}{14}{12}{9213}, $\delta_{J2000.0}$ =
+\decdms{28}{12}{12}{190}, and five hours of telescope time were allocated to
each epoch. The observing strategy (with cycles of two minutes on source and
one minute on the main calibrator) and the calibrators were the same as those
used for project BL128A.

\subsection{BM306}

Finally, a single observation of V773~Tau~A was obtained on September 2009
(epoch 27 in Table \ref{table_observations}) when the source was near
periastron. The data were collected using the High Sensitivity Array (HSA)
which combines the VLBA with the phased VLA, and the Arecibo, Green Bank, and
Effelsberg telescopes. However, for consistency with the other observations
presented in this paper, only the VLBA antennas were retained during the data
calibration. The phase center was at $\alpha_{J2000.0}$ =
\dechms{04}{14}{12}{9262}, $\delta_{J2000.0}$ = +\decdms{28}{12}{12}{102}. Six
hours of telescope time were allocated to the observation in this project, and
the observing strategy was the same as for projects BL128A and BL146.

\begin{figure}[!t]
  \centerline{\includegraphics[height=0.72\textwidth,angle=0]{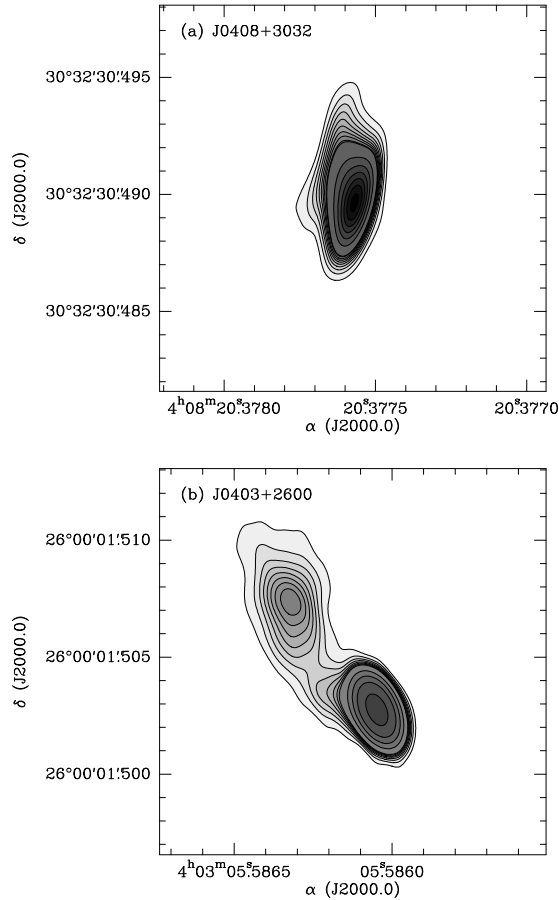}}
  \caption{Images of the two main calibrators used in this paper. (a)
    J0408$+$3032. The noise level in the image is $\sigma$ = 0.18 mJy, and the
    contours are at 5 to 50$\sigma$ by 5$\sigma$ and at 0.01 to 0.07 Jy by
    0.01 Jy. (b) J0403$+$260. The noise level in the image is $\sigma$ = 0.53
    mJy, and the contours are at 5 to 50$\sigma$ by 5$\sigma$ and at 0.05,
    0.1, 0.2, and 0.5 Jy.\label{fig_calibrators}}
\end{figure}

\begin{figure*}[!t]
  \centerline{\includegraphics[height=1\textwidth,angle=0]{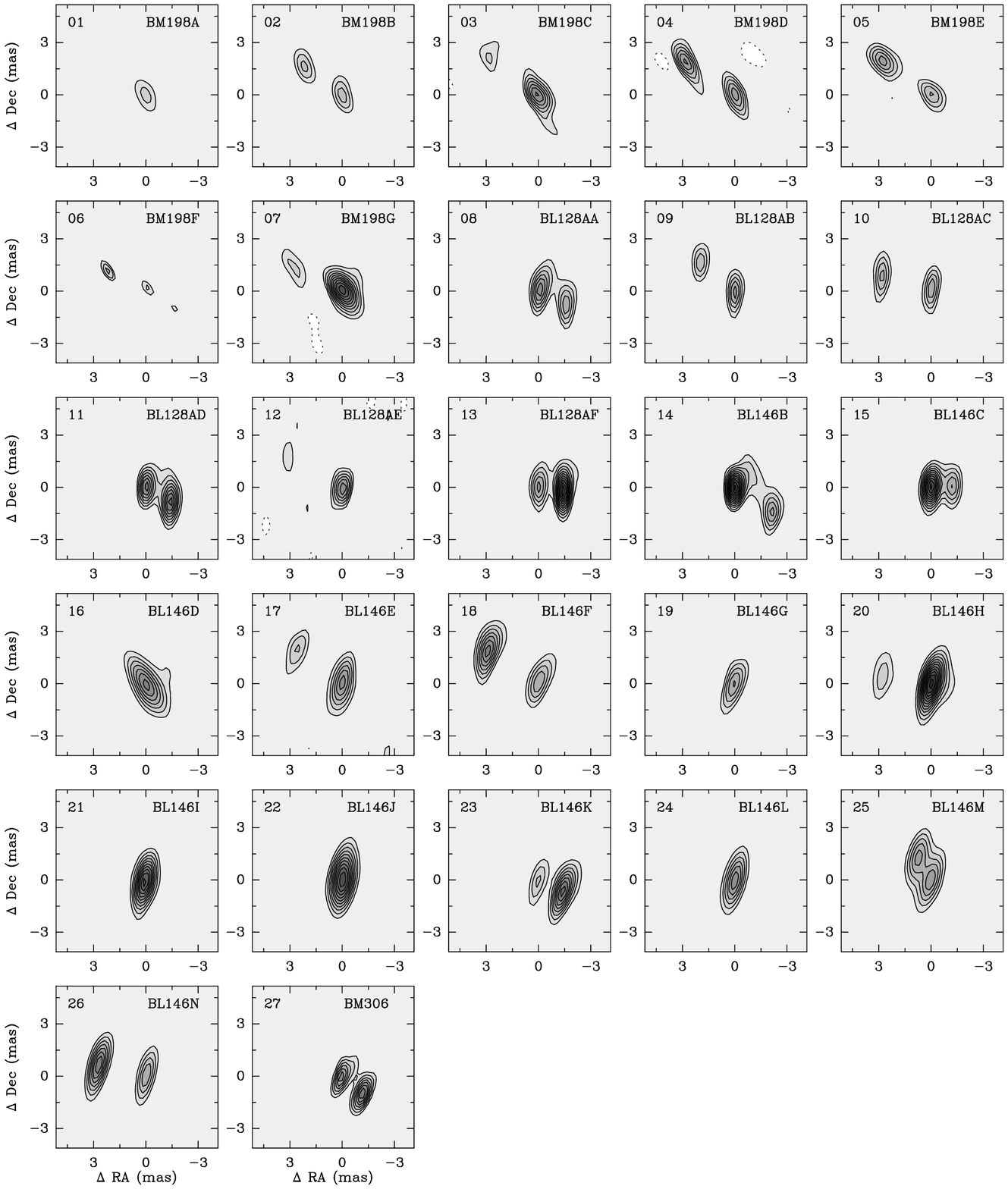}}
  \caption{VLBA images of V773~Tau~A at all 27 epochs. The contour levels have
    been adapted to each image. The sequence number, and code of the
    observation are indicated at the upper left and upper right corner of each
    panel respectively. The (0,0) center of each image is the position of the
    primary at the corresponding epoch.\label{fig_epochs}}
\end{figure*}

\bigskip

All four sets of data were edited and calibrated in a similar fashion using the
Astronomical Image Processing System (AIPS --Greisen 2003). The basic data
reduction followed the standard VLBA procedure for phase-referenced
observations. For the projects BL128A, BL146, and BM306, the data processing
included the multi-source calibration strategy described in detail in Torres
et al.\ (2007), and Torres (2010; Sections 4.2--4.5). This was not applied to
the data corresponding to the archival project BM198, which did not include
the necessary secondary calibrators scans. Once calibrated, the visibilities
were imaged with a pixel size of 50 $\mu$as applying weights intermediate
between natural and uniform (ROBUST = 0 in AIPS; Briggs 1995) to optimize the
compromise between angular resolution and sensitivity.

Obtaining polarimetric information is important to characterize the radio
emission mechanisms, but requires specific additional calibration scans (in
the present case, observations of 3C84 at the beginning, the middle, and the
end of each observing session) which were only obtained as part of project
BL146 (epochs 14--26 in Table \ref{table_observations}). These observations
are used to determine and correct for several instrumental and atmospheric
effects still corrupting the measured cross correlations after the standard
calibration procedure described above. Specifically, our polarization
calibration consisted of two steps. First, we calibrated the
cross-polarization delays using observations of the very strong calibrator
3C84 to remove differences in single and multiband delays for the right and
left handed components. Second, we computed instrumental polarization
corrections to calibrate the feed impurities (the so-called ``D-terms'') using
observations of 3C84 to determine the effective feed parameters for each
antenna and IF. Finally, the data were imaged in Stokes parameter $V$
(circular polarization) with the same pixel size used for Stokes $I$ (50
$\mu$as). An analysis of the issue of linear polarization will be deferred to
a forthcoming paper.

\bigskip

One important aspect concerning astrometry must be mentioned here. The
astrometric information derived from VLBI observations is contained in the
complex visibility phases calibrated against the main calibrator used during
the observations. As a consequence, it is very important (i) to use a main
calibrator as compact and structureless as possible, and (ii) to use the {\it
  same} calibrator for all observations. The last three sets of data used here
(BL128A, BL146, and BM306) were obtained using the same main calibrator
(J0408$+$3032), which is indeed very compact and largely structureless (Figure
\ref{fig_calibrators}a). Project BM198, however, was not originally intended
as an astrometric experiment and used a different calibrator (J0403$+$2600)
which has a well-resolved core-jet structure (Figure
\ref{fig_calibrators}b). This source was used as a secondary calibrator for
the other three projects, so the source positions derived from project BM198
can be indirectly registered to J0408$+$3032. The poorer quality
of the main calibrator and the indirect nature of the astrometry registration
imply, however, that the astrometry derived from project BM198 is somewhat
less accurate than that of the other three projects.

\begin{landscape}
\begin{deluxetable}{ccccccccccc}
\tablecaption{Measured source positions of the two components in V773 Tau, and orbit phase ($\phi$) of the system at all epochs.\label{table_positions_1}}
\tablehead{
\colhead{Epoch}&
\multicolumn{4}{c}{Primary (Aa)}&
\colhead{}&
\multicolumn{4}{c}{Secondary (Ab)}&
\colhead{$\phi$}\\%
[0.1cm]\cline{2-5}\cline{7-10}\\[-0.3cm]
{}&
{$\alpha$ [J2000.0]}&
{$\sigma_\alpha$}&
{$\delta$ [J2000.0]}&
{$\sigma_\delta$}&
{}&
{$\alpha$ [J2000.0]}&
{$\sigma_\alpha$}&
{$\delta$ [J2000.0]}&
{$\sigma_\delta$}&
{}\\%
{}&
\multicolumn{1}{c}{04$^h$14$^m$}&
{}&
\multicolumn{1}{c}{28$^\circ$12$'$}&
{}&
{}&
\multicolumn{1}{c}{04$^h$14$^m$}&
{}&
\multicolumn{1}{c}{28$^\circ$12$'$}&
{}&
{}}%
\startdata
01&\mmsec{12}{9190149}&\mmsec{0}{0000136}&\msec{12}{228446}&\msec{0}{000200}&&ND                &ND                 &ND               &ND              &0.32\\
02&\mmsec{12}{9190237}&\mmsec{0}{0000026}&\msec{12}{229524}&\msec{0}{000059}&&\mmsec{12}{9191887}&\mmsec{0}{0000027}&\msec{12}{231170}&\msec{0}{000062}&0.34\\
03&\mmsec{12}{9190008}&\mmsec{0}{0000036}&\msec{12}{229220}&\msec{0}{000077}&&\mmsec{12}{9192084}&\mmsec{0}{0000084}&\msec{12}{231515}&\msec{0}{000215}&0.34\\
04&\mmsec{12}{9190131}&\mmsec{0}{0000050}&\msec{12}{229236}&\msec{0}{000111}&&\mmsec{12}{9192214}&\mmsec{0}{0000056}&\msec{12}{231057}&\msec{0}{000124}&0.37\\
05&\mmsec{12}{9189953}&\mmsec{0}{0000036}&\msec{12}{228788}&\msec{0}{000063}&&\mmsec{12}{9192051}&\mmsec{0}{0000036}&\msec{12}{230700}&\msec{0}{000058}&0.39\\
06&\mmsec{12}{9190069}&\mmsec{0}{0000158}&\msec{12}{228791}&\msec{0}{000168}&&\mmsec{12}{9192015}&\mmsec{0}{0000093}&\msec{12}{229800}&\msec{0}{000256}&0.41\\
07&\mmsec{12}{9190049}&\mmsec{0}{0000031}&\msec{12}{228361}&\msec{0}{000059}&&\mmsec{12}{9192194}&\mmsec{0}{0000077}&\msec{12}{229758}&\msec{0}{000161}&0.43\\
08&\mmsec{12}{9215399}&\mmsec{0}{0000007}&\msec{12}{202276}&\msec{0}{000019}&&\mmsec{12}{9214304}&\mmsec{0}{0000007}&\msec{12}{201418}&\msec{0}{000024}&0.99\\
09&\mmsec{12}{9211595}&\mmsec{0}{0000011}&\msec{12}{196132}&\msec{0}{000037}&&\mmsec{12}{9213080}&\mmsec{0}{0000015}&\msec{12}{197895}&\msec{0}{000042}&0.32\\
10&\mmsec{12}{9207461}&\mmsec{0}{0000017}&\msec{12}{191012}&\msec{0}{000056}&&\mmsec{12}{9209591}&\mmsec{0}{0000023}&\msec{12}{191822}&\msec{0}{000076}&0.63\\
11&\mmsec{12}{9211497}&\mmsec{0}{0000007}&\msec{12}{186528}&\msec{0}{000020}&&\mmsec{12}{9210497}&\mmsec{0}{0000007}&\msec{12}{185703}&\msec{0}{000023}&0.01\\
12&\mmsec{12}{9218068}&\mmsec{0}{0000033}&\msec{12}{181160}&\msec{0}{000085}&&ND          &ND                &ND               &ND              &0.42\\
13&\mmsec{12}{9226934}&\mmsec{0}{0000011}&\msec{12}{178983}&\msec{0}{000035}&&\mmsec{12}{9225894}&\mmsec{0}{0000005}&\msec{12}{178760}&\msec{0}{000019}&0.08\\
14&\mmsec{12}{9240466}&\mmsec{0}{0000007}&\msec{12}{156238}&\msec{0}{000015}&&\mmsec{12}{9239000}&\mmsec{0}{0000018}&\msec{12}{155069}&\msec{0}{000045}&0.97\\
15&\mmsec{12}{9240254}&\mmsec{0}{0000004}&\msec{12}{155202}&\msec{0}{000011}&&\mmsec{12}{9239336}&\mmsec{0}{0000008}&\msec{12}{155274}&\msec{0}{000024}&0.08\\
16&NR                 &NR                &NR               &NR              &&NR              &NR                &NR               &NR              &0.22\\
17&\mmsec{12}{9239237}&\mmsec{0}{0000043}&\msec{12}{153290}&\msec{0}{000120}&&\mmsec{12}{9241176}&\mmsec{0}{0000068}&\msec{12}{155149}&\msec{0}{000176}&0.34\\
18&\mmsec{12}{9238939}&\mmsec{0}{0000018}&\msec{12}{153463}&\msec{0}{000042}&&\mmsec{12}{9241178}&\mmsec{0}{0000012}&\msec{12}{155184}&\msec{0}{000032}&0.43\\
19&\mmsec{12}{9238880}&\mmsec{0}{0000036}&\msec{12}{153258}&\msec{0}{000097}&&ND            &ND                &ND               &ND              &0.53\\
20&\mmsec{12}{9239130}&\mmsec{0}{0000010}&\msec{12}{152788}&\msec{0}{000027}&&\mmsec{12}{9241227}&\mmsec{0}{0000044}&\msec{12}{153447}&\msec{0}{000112}&0.65\\
21&NR                 &NR                &NR               &NR              &&NR              &NR                &NR               &NR              &0.77\\
22&NR                 &NR                &NR               &NR              &&NR              &NR                &NR               &NR              &0.88\\
23&\mmsec{12}{9240573}&\mmsec{0}{0000011}&\msec{12}{152076}&\msec{0}{000036}&&\mmsec{12}{9239482}&\mmsec{0}{0000006}&\msec{12}{151553}&\msec{0}{000016}&0.04\\
24&NR                 &NR                &NR               &NR              &&NR              &NR                &NR               &NR              &0.16\\
25&\mmsec{12}{9239225}&\mmsec{0}{0000012}&\msec{12}{150350}&\msec{0}{000034}&&\mmsec{12}{9239774}&\mmsec{0}{0000010}&\msec{12}{151708}&\msec{0}{000029}&0.24\\
26&\mmsec{12}{9237209}&\mmsec{0}{0000006}&\msec{12}{149354}&\msec{0}{000022}&&\mmsec{12}{9239280}&\mmsec{0}{0000005}&\msec{12}{149924}&\msec{0}{000014}&0.64\\
27&\mmsec{12}{9272872}&\mmsec{0}{0000012}&\msec{12}{100333}&\msec{0}{000026}&&\mmsec{12}{9272011}&\mmsec{0}{0000009}&\msec{12}{099345}&\msec{0}{000021}&0.95\\
\enddata
\footnotetext{NR=not resolved, ND=not detected}
\end{deluxetable}

\end{landscape}

\section{Astrometry}

The images of V773~Tau~A obtained at all 27 epochs are shown in Figure
\ref{fig_epochs}. A well-resolved double source morphology (corresponding to
the two stars in V773~Tau~A) is detected at 20 of the 27 epochs, while a
single source is seen in the remaining seven epochs. The source positions
(Table \ref{table_positions_1}) were determined using a two-dimensional Gaussian
fitting procedure (task JMFIT in AIPS); when the source was double, both
components were fitted simultaneously. JMFIT provides an estimate of the
errors on the source position based on the expectation for ideal
interferometer data. However, in typical VLBI observations where the main
calibrator is located a few degrees from the astronomical target (such as
those considered here), remaining systematic phase calibration errors normally
dominate the error budget, so the errors provided by JMFIT underestimate the
true uncertainties. We will come back to this point later.

\subsection{Relative Astrometry \& Orbit Modeling}

In this section, we will concentrate on the 20 epochs when the source was
double, and analyze the relative position of the two sub-components. As
established by P1996 and B2007, these double source observations reflect the
relative astrometry of the V773~Tau~A components,\footnote{As will become
  apparent momentarily, the emission is indeed associated with the two stars,
  but in some cases, it is not exactly {\em coincident} with them.} and as
such can be used to assess and update the physical orbit obtained by
B2007. The angular separations ($\Delta\alpha$, $\Delta\delta$) between the
primary and the secondary are given in Table \ref{table_separations} for the
20 epochs when the source is double.  The uncertainties on ($\Delta\alpha$,
$\Delta\delta$) quoted in Table \ref{table_separations} are based on the
errors delivered by JMFIT, and are almost certainly underestimated (see
above).

\begin{figure}[!t]
  \centerline{\includegraphics[height=0.45\textwidth,angle=0]{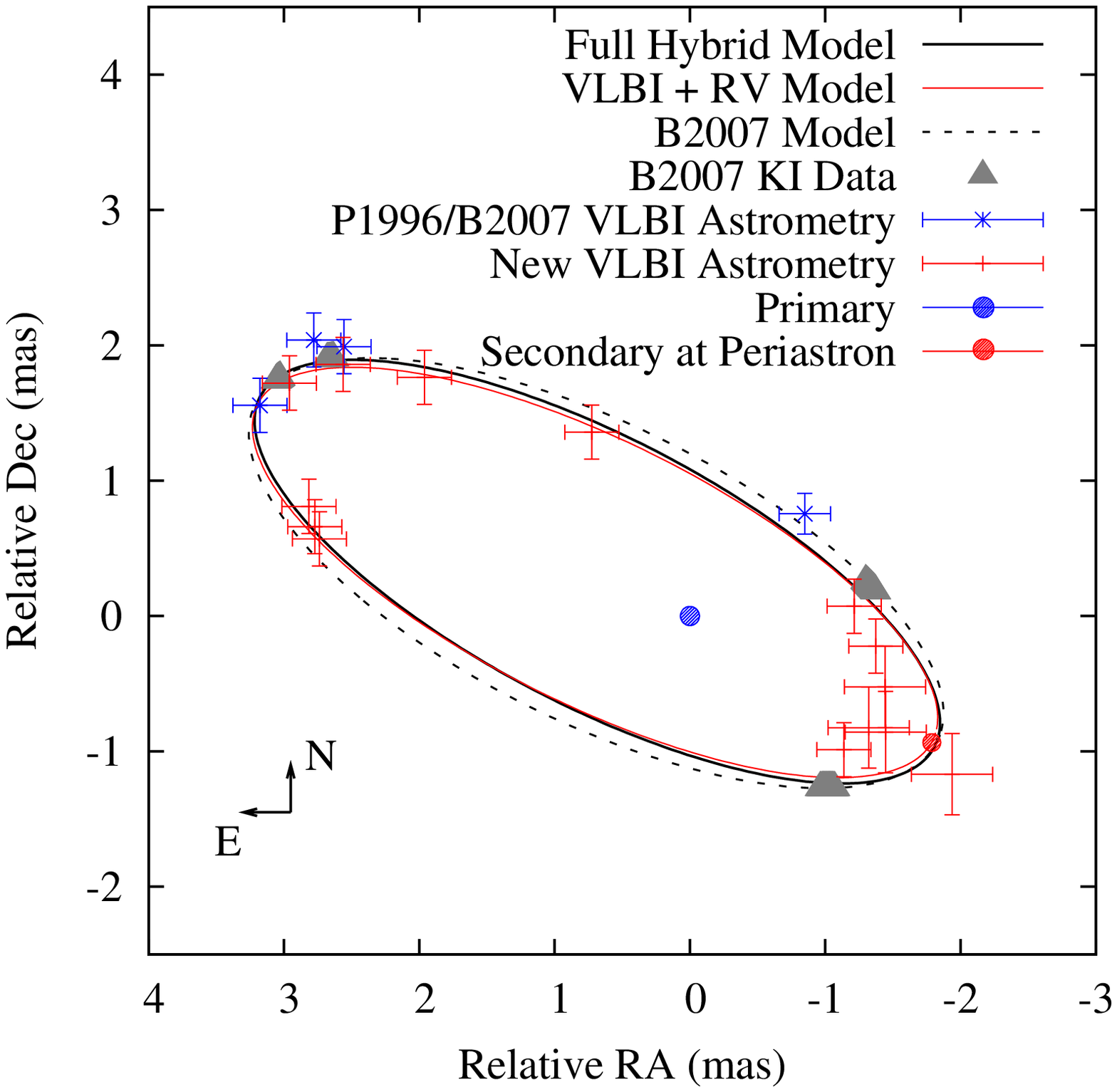}}
  \caption{VLBI Relative Astrometry and Orbit Models of V773~Tau~A.  Here VLBI
    relative astrometry from P1996, B2007 and this work is shown with orbit
    model renderings for the B2007 orbit and two models derived here (the
    ``VLBI + RV'' and ``Full Hybrid'' models described in the text). The
    primary (Aa) component is shown at the origin and the secondary (Ab) at
    periastron. The apparent sizes of the V773~Tau~A components are estimated
    by SED modeling (B2007) and rendered to scale.\label{fig_orbit}}
\end{figure}

Figure \ref{fig_orbit} shows all available relative astrometry data on
V773~Tau~A (from P1996, B2007, and this work), along with several orbit models
including that from B2007 (dashed line). Clearly there is general agreement
between the old and new VLBI relative astrometry on V773~Tau~A, and the a
priori B2007 orbit model. A close inspection of the VLBI separations and the
B2007 orbit model in Figure \ref{fig_orbit}, however, shows an interesting
trend in the VLBI astrometry. At most orbit phases the VLBI-derived
separations and B2007 orbit model are in good agreement. But in most
observations near periastron (secondary south-west of the primary) the VLBI
separations appear systematically smaller (i.e.\ secondary nearer the primary)
than predicted by the B2007 orbit. This suggests one of two possibilities:
there is a possible bias in the B2007 orbit solution, or a possible bias in
the VLBI-derived astrometry near periastron (where there is known enhancement
in the radio flaring; Massi et al.\ 2002, 2006).

{\scriptsize
\begin{deluxetable}{crrrrr}
\tablecaption{Measured separations and orbit phases.\label{table_separations}}
\tablehead{
\colhead{Epoch}&
\colhead{$\Delta \alpha$}&
\colhead{$\sigma_{\Delta\alpha}$}&
\colhead{$\Delta \delta$}&
\colhead{$\sigma_{\Delta\delta}$}&
\colhead{$\phi$}\\%
{}&
{[mas]}&
{}&
{[mas]}&
{}&
{}}%
\startdata
01 & --    & --   & --    & --   & 0.32  \\
02 &  2.18 & 0.04 &  1.65 & 0.06 & 0.34  \\
03 &  2.74 & 0.11 &  2.29 & 0.22 & 0.34  \\
04 &  2.75 & 0.08 &  1.82 & 0.13 & 0.37  \\
05 &  2.77 & 0.06 &  1.91 & 0.07 & 0.39  \\
06 &  2.90 & 0.15 &  3.43 & 0.23 & 0.41  \\
07 &  2.83 & 0.11 &  1.40 & 0.16 & 0.43  \\
08 & -1.45 & 0.01 & -0.86 & 0.03 & 0.99  \\
09 &  1.96 & 0.02 &  1.76 & 0.06 & 0.32  \\
10 &  2.82 & 0.04 &  0.81 & 0.09 & 0.63  \\
11 & -1.32 & 0.01 & -0.83 & 0.03 & 0.01  \\
12 & --    & --   & --    & --   & 0.42  \\
13 & -1.37 & 0.02 & -0.22 & 0.04 & 0.08  \\
14 & -1.94 & 0.02 & -1.17 & 0.05 & 0.97  \\
15 & -1.21 & 0.01 &  0.07 & 0.03 & 0.08  \\
16 & --    & --   & --    & --   & 0.22  \\
17 &  2.56 & 0.11 &  1.86 & 0.21 & 0.34  \\
18 &  2.96 & 0.03 &  1.72 & 0.05 & 0.43  \\
19 & --    & --   & --    & --   & 0.53  \\
20 &  2.77 & 0.06 &  0.66 & 0.12 & 0.65  \\
21 & --    & --   & --    & --   & 0.77  \\
22 & --    & --   & --    & --   & 0.88  \\
23 & -1.44 & 0.02 & -0.52 & 0.04 & 0.04  \\
24 & --    & --   & --    & --   & 0.16  \\
25 &  3.37 & 0.02 &  2.05 & 0.04 & 0.24  \\
26 &  2.74 & 0.01 &  0.57 & 0.03 & 0.64  \\
27 & -1.14 & 0.02 & -0.99 & 0.03 & 0.95  \\
\enddata
\footnotetext[1]{For completeness on the orbit phases, the epochs when only one source is detected are also given.}
\end{deluxetable}
}

To investigate this issue we considered two different prescriptions for
integrating the new VLBI data into the orbit modeling. In the first ``VLBI +
RV'' model we considered the possibility that the Keck Interferometer (KI)
visibilities were the source of a possible orbit bias in B2007, and used only
the (old and new) VLBI-derived relative astrometry and double-lined RV from
B2007 to derive an orbit model. In the second ``Full Hybrid'' model we used
all available data (VLBI astrometry, KI visibilities, and RV), but assume that
the VLBI separations near periastron were biased by the enhanced flaring
activity (e.g.\ if the enhanced flares preferentially occur between the two
stars), assigning these points lower weight (specifically a factor of 50\%
larger 1-sigma error per axis for orbit phases within 10\% of
periastron).\footnote{Our methods for orbit modeling with heterogenous RV and
  astrometric/visibility datasets are described in Boden et al.\ (2000) and
  are not repeated here. In all the orbit modeling we have weighted all VLBI
  astrometry consistently with the weighting derived in B2007, at 0.2 mas per
  axis 1-sigma, except as noted for points near periastron in the ``Full
  Hybrid'' model.} We find that both new orbit solutions are in excellent
agreement with the original B2007 model, and renderings of both these orbit
models are included in Figure \ref{fig_orbit}. It is clear that the KI
visibility data is reliable at its stated uncertainties, and the most
plausible hypothesis is that the VLBI relative astrometry near periastron
contains biases associated with the enhanced flaring activity documented by
Massi et al.\ (2002, 2006). This astrometric bias may be naturally explained
in the helmet streamer scheme recently proposed by Massi et al.\
(2008). However, a more detailed analysis, which we defer to a forthcoming
paper, will be necessary to understand better the origin of the apparent
offset between the radio sources and the underlying stars.


For our final A-subsystem orbit estimation we conducted a joint orbit
modeling with the A-B astrometry and RV datasets presented in the B2011
companion paper.  The basis of the joint modeling recognizes that RV
motions of the A-component stars also include the motion of the
A-barycenter -- moving under the gravitational influence of the other
V773~Tau components, and dominated by the A-B orbital motion.  B2011
describes the joint orbit modeling of both the A-subsystem and the A-B
orbit based on the combined VLBA relative astrometry, KI visibility,
high-resolution imaging, and expanded RV datasets. Table \ref{table_orbital_parameters} gives a direct comparison between
the orbit models from B2007 and the ``Joint Solution'' model derived
in B2011 based in part on data presented here; in all cases the
orbital parameters between the two models (which share a significant
amount of underlying data) are in excellent statistical agreement. For
all subsequent analysis we adopt this ``Joint Solution'' model for the
updated V773~Tau~A orbit.

{\scriptsize
\begin{deluxetable}{lr@{$\pm$}lr@{$\pm$}l}
\tablecaption{V773~Tau~A Orbital Parameters.\label{table_orbital_parameters}}
\tablehead{
\colhead{Orbital Parameter} &
\multicolumn{2}{c}{B2007} &
\multicolumn{2}{c}{``Full Hybrid''} }%
\startdata
Period [days] \dotfill & 51.1039 & 0.0021 & 51.1033 & 0.0018 \\
\textit{T$_o$} [MJD] \dotfill & 53059.73 & 0.33 & 53059.75 & 0.28 \\
\textit{e} \dotfill & 0.2717 & 0.0085 & 0.2713 & 0.0066 \\
\textit{K$_A$} [\kmps] \dotfill & 35.90 & 0.53 & 35.72 & 0.46 \\%
\textit{K$_B$} [\kmps] \dotfill & 41.5 & 1.4 & 42.9 & 1.3 \\%
$\gamma$ [\kmps] \dotfill & 0.02 & 0.32 & 0.03 & 0.32 \\
$\omega_A$ [deg] \dotfill & 4.6 & 2.4 & 5.6 & 2.2 \\
$\Omega$ [deg] \dotfill & 63.5 & 1.7 & 62.4 & 1.1 \\
\textit{i} [deg] \dotfill & 66.0 & 2.4 & 68.5 & 1.6 \\
\textit{a} [mas] \dotfill & 2.811 & 0.047 & 2.809 & 0.033  \\
\enddata
\end{deluxetable}
}


Given the agreement between the ``Joint Solution'' model and the B2007
orbit model, the updated physical parameters are highly consistent
with B2007 estimates. Component dynamical masses resulting from this
orbit model 1.59 $\pm$ 0.12 \Msun\ and 1.323 $\pm$ 0.079 \Msun, for the
primary and the secondary respectively -- consistent with the B2007
estimates to 3\% and 0.4-sigma. Similarly, the system distance
estimate derived from the updated orbit model is 135.7 $\pm$ 3.2 pc,
again highly consistent with the value from B2007. As a final note, we
should stress that the present model must still be considered
preliminary as astrometric and RV observations to assess the
gravitational effect of the two other members of the system
(V773~Tau~B and C, reported in B2011) are ongoing.

\subsection{Absolute Astrometry \& Parallax determination}

In addition to their separation, the VLBA provides the absolute position of
each of the radio sources (Table \ref{table_positions_1})\footnote{Strictly, the
  VLBA provides the angular offset between the source and the main
  calibrator. However, that calibrator is a distant quasar which can be
  assumed to be fixed, and whose coordinates are measured relative to the
  International Celestial Reference Frame. Thus, the positions delivered by the VLBA are very nearly
  absolute.}. Once the orbital motion of the system has been determined, it
can be subtracted from the measured positions to obtain the coordinates of the
barycenter of the system. If the measured separation between the primary and
the secondary is $\mathbf r$, then the separation $\mathbf{r}_1$ between the
primary to the barycenter is given by

\begin{equation}
  \mathbf{r}_1 = {m_s \over m_s + m_p} \mathbf{r},
\end{equation}

\begin{figure}[!t]
  \centerline{\includegraphics[width=0.5\textwidth,angle=270]{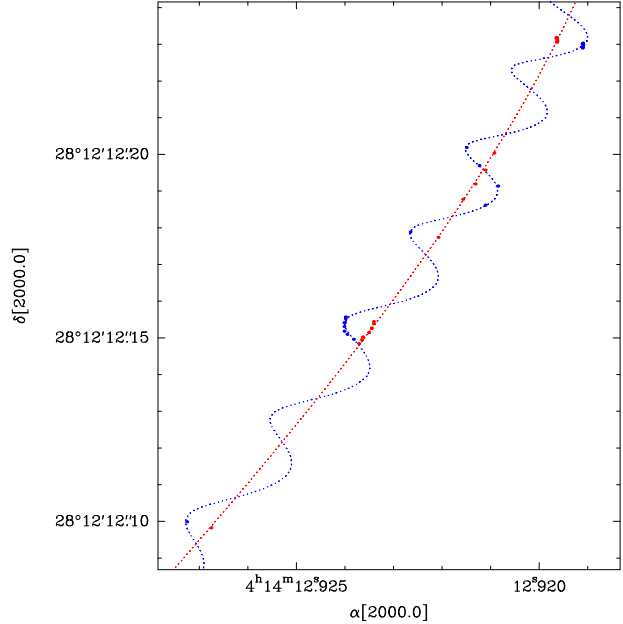}}
  \caption{Measured barycenter positions (blue symbols) and best astrometric
    fit (blue dotted curve) for V773~Tau A. In both cases, the observed positions are 
    shown as ellipses, whose sizes are the error bars. The red symbols and curve
    show the same positions but with the parallactic contribution
    removed. Note the clear curvature of the trajectory of the barycenter of
    V773~Tau~A.\label{fig_parallax}}
\end{figure}

\noindent
where $m_p$ and $m_s$ are the masses of the primary and secondary,
respectively, deduced from the orbital fit. Clearly, this strategy can only be
applied directly to our 20 observations where the system is found to be
double. We did not attempt to correct for the displacement of the radio
emission from the stars themselves near periastron since it is not clear 
whether this effect preferentially affects one specific star or both equally. 

A fit to the resulting 20 barycenter positions (given explicitly in Table 3) with a 
combination of a trigonometric parallax and a uniform proper motion provides a 
very poor agreement with the observations. This is not particularly surprising since
V773~Tau~A will be affected by the gravitational influence of the B component
on the timescale of our dataset. In particular, recent infrared observations
(B2011) have shown that the mass of component B is 2.35 $\pm$ 0.67 \Msun\ (i.e.\
comparable to that of component A), while the semi-major axis and period of
the A-B orbit is approximately 15 AU and 26 yr, respectively. Thus, our
observations cover a time span corresponding to about a fifth of the total
orbital period, and the acceleration of the barycenter of V773~Tau~A is
expected to be significant. Indeed, a fit with a uniformly accelerated proper
motion provides an excellent description of the data (Figure
\ref{fig_parallax}). The astrometric elements corresponding to that fit are:

\begin{eqnarray*}
  \alpha_{J2006.97} & = & 04^{{\rm h}}14^{{\rm m}}12\rlap.{^{\rm s}}922447\pm 0.000010\\
  \delta_{J2006.97} & = &\ 28^{\circ}12^{'}12\rlap.{''}170682 \pm 0.000097\\
  \mu_{\alpha} \cos{\delta}_{J2006.97}&=&17.092 \pm 0.077\ {\rm mas\ yr}^{-1}\\
  \mu_{\delta~{J2006.97}}&=&-24.030 \pm 0.053\ {\rm mas\ yr}^{-1}\\
  a_\alpha \cos{\delta}&=&2.60 \pm 0.60\ {\rm mas\ yr}^{-2}\\
  a_\delta&=&-1.51 \pm 0.52\ {\rm mas\ yr}^{-2}\\
  \pi&=&7.70 \pm0.19\ {\rm mas}.
\end{eqnarray*}

The reference epoch (2006.97) corresponds to the median epoch of the
observations used in the astrometric fit. The post-fit r.m.s.\ is 0.34 mas in
right ascension, and 0.29 mas in declination, in good agreement with our
assumption that the errors on the separation between the sources are of the
order of 0.2--0.3 mas per axis (see Section 3.1). The acceleration found here
is in excellent agreement with the expected acceleration of the barycenter of
V773~Tau~A due to the gravitational pull of component V773~Tau~B. In the
companion paper, B2011 report on multi-epoch near-infrared observations of and
orbit modeling for the A-B subsystem. The mean A-barycenter acceleration 
estimated from that orbit model during the time span covered by our VLBA 
observations is $a_\alpha \cos{\delta}$ = 2.21 $\pm$ 0.55 ${\rm mas\ yr}^{-2}$; 
$a_\delta = -1.24\pm 0.35 \ {\rm mas\  yr}^{-2}$, both well within 1$\sigma$ of the VLBA 
determination. It is important to stress that the two determinations of the
acceleration are entirely independent --one is based on absolute astrometry of
radio observations, while the other rests on relative astrometry of
near-infrared data. Thus, the concordance of the two results lends strong
support to both analyses.

The distance corresponding to the trigonometric parallax obtained from our fit
is 129.9 $\pm$ 3.2 pc. This is in agreement with the --almost entirely
independent-- distance obtained from the orbit model at the 4.5\%, or
1$\sigma$ level.\footnote{Here, and in the rest of the paper, we test the
  consistency of results by examining the number of sigmas required to obtain
  overlapping error bars. In the present case, the 1$\sigma$ error bar on the
  orbit distance estimate is 132.5 pc $< d <$ 138.9 pc, while for the
  trigonometric parallax distance, it is 126.7 pc $< d <$ 133.1 pc). Since
  they overlap, we consider the results consistent within 1$\sigma$.} This
very good level of agreement validates both methods, and demonstrates that the
quoted errors represent the true accuracy (and not just the precision) of the
results. By taking the mean of the two values, we obtain our favored estimate
of the distance to V773~Tau~A: $d$ = 132.8 $\pm$ 2.3 pc.

Our distance determination is only very marginally consistent with the result
($d$ = 148.4$_{-5.3}^{+5.7}$ pc) obtained by Lestrade et al.\ (1999) using
multi-epoch VLBI observations similar to those presented here. The trustworthiness
of our astrometry is demonstrated by the concordance of our parallax,
acceleration, and proper motion (see below) with those of entirely 
independent infrared observations. Lestrade et al.\ (1999), on the other 
hand, did not take into account the binarity of the A subsystem nor the effect 
of the B component in their astrometric modeling. As a consequence, the
mathematical function used in their fit did not provide an adequate 
description of the modeled data, and the quoted errors on the fitted 
parameters do not properly reflect  actual uncertainties. To reconcile
the results of Lestrade et al.\ (1999) with the conclusions of our astrometric
analyses, the errors quoted by Lestrade et al.\ (1999) must be multiplied
by a factor of about 3. We conclude that the present work improves the
actual uncertainty on the distance to V773 Tau by nearly one order of magnitude
from about 15--20 pc down to 2.4 pc) over the results by Lestrade et al.\ 
(1999).

Another apparent discrepancy with the results of Lestrade et al.\ (1999) must
be briefly mentioned here. Lestrade et al.\ (1999) reported proper motions for
V773~Tau~A of $\mu_\alpha \cos \delta$ = 0.42 $\pm$ 0.29 mas yr$^{-1}$;
$\mu_\delta$ = --23.25 $\pm$ 0.25 mas yr$^{-1}$. Those figures are similar to
those reported in the Hipparcos catalog: $\mu_\alpha \cos \delta$ = 0.65 $\pm$
2.61 mas yr$^{-1}$; $\mu_\delta$ = --24.89 $\pm$ 1.81 mas yr$^{-1}$ (Perryman
et al.\ 1997), but are very different from those found here especially in
right ascension. The difference is $\Delta (\mu_\alpha \cos \delta)$ = +16.7
$\pm$ 0.3 mas yr$^{-1}$; $\Delta(\mu_ \delta)$ = --0.8 $\pm$ 0.3 mas
yr$^{-1}$. The difference once again reflects the non-uniformity of the
barycentric motion of V773~Tau~A due to the gravitational effect of
V773~Tau~B. While the mean epoch of the observations reported by Lestrade et
al.\ (1999) and of the Hipparcos satellite are very similar (1993.88, and
1991.25, respectively), our observations are significantly more recent
(2006.97). According to the A-B orbit published in B2011, the instantaneous
proper motion of the A-barycenter of V773~Tau (measured relative to the
barycenter of the entire system) was $\mu_\alpha \cos \delta$ = --8.85 mas
yr$^{-1}$; $\mu_\delta$ = +0.42 mas yr$^{-1}$ at epoch 1993.88, and
$\mu_\alpha \cos \delta$ = +8.81 mas yr$^{-1}$; $\mu_\delta$ = +0.40 mas
yr$^{-1}$ at epoch 2006.97. Thus, the difference in proper motion between
1993.88 and 2006.97 according to the A-B orbital fit is $\Delta (\mu_\alpha
\cos \delta)$ = +17.66 mas yr$^{-1}$; $\Delta(\mu_ \delta)$ = --0.02 mas
yr$^{-1}$, in very good agreement with the measured values. Indeed, 
knowing the relative expected proper motion between the barycenter of 
V773~Tau~A and the barycenter of V773~Tau as a whole, we can estimate 
the proper motion of the barycenter of V773~Tau as a whole. We obtain 
$\mu_\alpha \cos \delta$ = +8.3 $\pm$ 0.5 mas yr$^{-1}$; $\Delta(\mu_ \delta)$ 
= --23.6 $\pm$ 0.5 mas yr$^{-1}$.

\smallskip

We mentioned earlier that in 7 of our 27 observations, the source was
single. With the help of the astrometric fits, it is possible to determine the
state of the system at those epochs. In three of the cases (epochs 1, 12, and
19), the single detected source appears to be the primary of the system. In
all three cases, the source was weak ($\sim$ 1 mJy), and the system was near
apastron. Evidently, the secondary had faded below our detection limit at
these epochs. In the remaining four observations (epochs 16, 21, 22, and 24),
the single detected source is bright (at least several mJy), and its position
is intermediate between the expected positions of the primary and the
secondary. In all cases, the system was at an orbit phase of about 0.2 before
or after periastron. In this situation, the two stars are located in
projection almost exactly north-south of each other, with a projected
separation of about 1 mas (see Figure \ref{fig_orbit}). Since our resolution
in the north-south direction is about 2 mas, we do not expect to resolve the
two stars in this situation, but instead to detect a single source slightly
elongated in the north-south direction. This is in fact what happens (see
Figure \ref{fig_epochs}). Indeed, the mean deconvolved FWHM size of the
emission for these four epochs is 1.33 mas, 50\% larger than the corresponding
figure (0.89 mas) when the source is double or single but near apastron.

\subsection{The distance to the L1495 region of Taurus}

In projection, V773~Tau is located toward the dark cloud Lynds~1495 in the
central region of Taurus. Two other young stars with recent VLBA-based
parallax measurements (Hubble~4 and HDE~283572; Torres et al.\ 2007) are
located in the same portion of Taurus. Interestingly, they appear to be at
very similar distances (132.8 $\pm$ 0.5 pc for Hubble 4 and 128.5 $\pm$ 0.6 pc
for HDE~283572, against 132.8 $\pm$ 2.3 pc for V773~Tau). The proper motions
of Hubble~4 and HDE~283572 ($\mu_\alpha \cos \delta$ = +4.30 $\pm$ 0.05 mas
yr$^{-1}$; $\mu_ \delta$ = --28.9 $\pm$ 0.3 mas yr$^{-1}$, and $\mu_\alpha
\cos \delta$ = +8.88 $\pm$ 0.06 mas yr$^{-1}$; $\mu_ \delta$ = --26.6 $\pm$
0.1 mas yr$^{-1}$, respectively --Torres et al.\ 2007) are similar to one
another, and with that of the barycenter of V773~Tau ($\mu_\alpha \cos \delta$ 
= +8.3 $\pm$ 0.5 mas yr$^{-1}$; $\mu_ \delta$ = --23.6 $\pm$ 0.5 mas
yr$^{-1}$; Section 3.2). The radial velocities of the three stars are also
similar: 15.0 $\pm$ 1.7 km s$^{-1}$ for Hubble 4 (Hartmann et al.\ 1986), 15.0 
$\pm$ 1.5 km s$^{-1}$ for HDE~283572 (Walter et al.\ 1988), and 16.38 $\pm$
0.52 km s$^{-1}$ for V773 Tau (B2011). Finally, the extinction measured toward 
the V773 Tau system (B2011) is consistent with extinction estimated by Schmalzl
et al.\ 2010) for that portion of L1495. Thus, Hubble~4, HDE~283572, and 
V773~Tau are not only located in the same portion of Taurus, they are also 
at similar distances, and share the same kinematics: they most likely belong
to a common spatio-kinematical sub-group within the Taurus complex.
We note that in a previous paper of this series (Torres et   al.\ 2007), we 
reached an opposite conclusion because we used the distance to V773~Tau 
and proper motions given by Lestrade et al.\ (1999). The present work shows 
that the latter distance was likely overestimated by about 15 pc,  and that the 
proper motions have to be corrected for the A-B orbit. We argue that the weighted 
mean of our three parallax measurements provides a good estimate of the mean 
distance to Lynds~1495, and that their dispersion is a good measure of the 
depth of the associated stellar population.  We conclude that Lynds~1495 can 
be taken to be at 131.4 $\pm$ 2.4 pc.

\begin{figure}[!t]
  \centerline{\includegraphics[width=0.5\textwidth,angle=0]{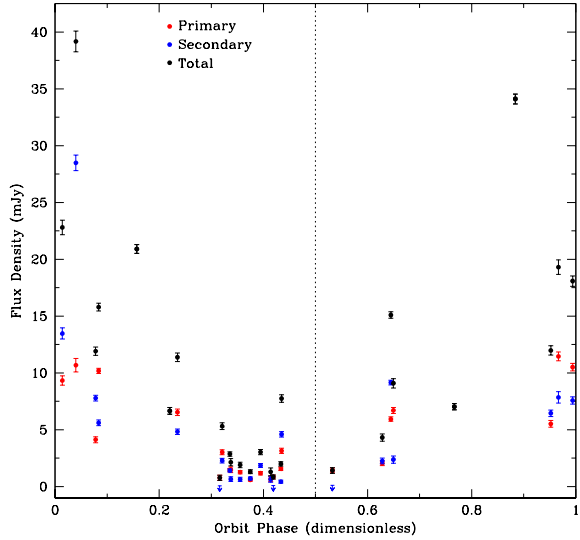}}
  \caption{Measured fluxes of V773~Tau as a function of orbit phase (0 and 1
    mean periastron, whereas 0.5 corresponds to apastron). The total flux of
    the system as well as the individual fluxes of the sub-components are
    shown in different colors.\label{fig_flux}}
\end{figure}

\begin{deluxetable}{crrrrrrrrrc}
\tablecaption{Source fluxes and brightness temperatures.\label{table_fluxes}}
\tablehead{
{}&
\multicolumn{2}{c}{Primary (Aa)}&{}&
\multicolumn{2}{c}{Secondary (Ab)}&{}&
\multicolumn{2}{c}{}&{}&{}\\%
[0.1cm]\cline{2-3}\cline{5-6}\\[-0.3cm]
{Epoch\footnotemark[1]}&
\multicolumn{1}{c}{$f_{\nu}$}&\multicolumn{1}{c}{$T_{b}$}&{}&
\multicolumn{1}{c}{$f_{\nu}$\footnotemark[2]}&\multicolumn{1}{c}{$T_{b}$}&{}&
\multicolumn{1}{c}{$f_{\nu_{Aa}}+f_{\nu_{Ab}}$}&{$f_{\nu_{Ab}}/f_{\nu_{Aa}}$}&{r.m.s.}\\%
{}&
\multicolumn{1}{c}{[mJy]}&\multicolumn{1}{c}{[$10^6$ K]}&{}&
\multicolumn{1}{c}{[mJy]}&\multicolumn{1}{c}{[$10^6$ K]}&{}&
\multicolumn{1}{c}{[mJy]}&{}&[mJy]}%
\startdata
01 &  0.33$\pm$0.07 &   3.2$\pm$0.7 &&        $<$0.22 &        $<$2.1 &&  0.33$\pm$0.10 & $<$0.66 & 0.07 \\
02 &  1.43$\pm$0.08 &  14.2$\pm$0.8 &&  1.33$\pm$0.08 &  13.2$\pm$0.8 &&  2.76$\pm$0.11 &    0.93 & 0.08 \\
03 &  0.99$\pm$0.07 &   9.6$\pm$0.7 &&  0.35$\pm$0.07 &   3.4$\pm$0.7 &&  1.35$\pm$0.10 &    0.36 & 0.07 \\
04 &  0.70$\pm$0.07 &   5.9$\pm$0.6 &&  0.69$\pm$0.07 &   5.9$\pm$0.6 &&  1.40$\pm$0.10 &    0.99 & 0.07 \\
05 &  1.05$\pm$0.08 &  11.1$\pm$0.8 &&  1.23$\pm$0.08 &  13.0$\pm$0.8 &&  2.27$\pm$0.11 &    1.17 & 0.08 \\
06 &  0.50$\pm$0.08 &   4.9$\pm$0.8 &&  0.34$\pm$0.08 &   3.4$\pm$0.8 &&  0.84$\pm$0.12 &    0.69 & 0.09 \\
07 &  1.15$\pm$0.07 &  10.6$\pm$0.6 &&  0.36$\pm$0.07 &   3.3$\pm$0.6 &&  1.51$\pm$0.09 &    0.31 & 0.07 \\
08 &  7.65$\pm$0.15 &  89.5$\pm$1.8 &&  5.96$\pm$0.16 &  69.7$\pm$1.8 && 13.61$\pm$0.22 &    0.78 & 0.15 \\
09 &  2.56$\pm$0.11 &  32.0$\pm$1.3 &&  1.99$\pm$0.11 &  24.9$\pm$1.4 &&  4.55$\pm$0.15 &    0.79 & 0.11 \\
10 &  1.77$\pm$0.11 &  19.7$\pm$1.2 &&  1.49$\pm$0.11 &  16.6$\pm$1.2 &&  3.25$\pm$0.15 &    0.84 & 0.10 \\
11 &  8.66$\pm$0.22 & 110.1$\pm$2.9 &&  9.18$\pm$0.22 & 116.7$\pm$2.8 && 17.84$\pm$0.31 &    1.06 & 0.21 \\
12 &  0.80$\pm$0.09 &   9.7$\pm$1.1 &&        $<$0.28 &        $<$3.4 &&  0.80$\pm$0.13 & $<$0.35 & 0.09 \\
13 &  3.23$\pm$0.13 &  40.8$\pm$1.6 &&  6.23$\pm$0.13 &  78.7$\pm$1.6 &&  9.46$\pm$0.18 &    1.93 & 0.12 \\
14 &  9.56$\pm$0.19 & 114.7$\pm$2.3 &&  4.13$\pm$0.19 &  49.5$\pm$2.2 && 13.69$\pm$0.27 &    0.43 & 0.20 \\
15 &  9.23$\pm$0.13 & 106.2$\pm$1.5 &&  4.51$\pm$0.13 &  51.8$\pm$1.4 && 13.74$\pm$0.18 &    0.49 & 0.12 \\
16 & ---            & ---           && ---            & ---           &&  3.98$\pm$0.12 & ---     & 0.12 \\
17 &  1.09$\pm$0.12 &   8.3$\pm$0.9 &&  0.66$\pm$0.12 &   5.0$\pm$0.9 &&  1.75$\pm$0.16 &    0.60 & 0.12 \\
18 &  3.01$\pm$0.13 &  21.2$\pm$0.9 &&  4.19$\pm$0.13 &  29.5$\pm$0.9 &&  7.20$\pm$0.18 &    1.39 & 0.12 \\
19 &  1.27$\pm$0.12 &  10.7$\pm$1.0 &&        $<$0.36 &        $<$3.0 &&  1.27$\pm$0.17 & $<$0.28 & 0.12 \\
20 &  5.31$\pm$0.13 &  40.1$\pm$1.0 &&  1.37$\pm$0.13 &  10.4$\pm$0.9 &&  6.68$\pm$0.18 &    0.26 & 0.13 \\
21 & ---            & ---           && ---            & ---           &&  5.63$\pm$0.14 & ---     & 0.14 \\
22 & ---            & ---           && ---            & ---           && 24.01$\pm$0.20 & ---     & 0.21 \\
23 & 10.66$\pm$0.35 &  89.1$\pm$2.9 && 22.68$\pm$0.34 & 189.8$\pm$2.9 && 33.34$\pm$0.49 &    2.13 & 0.34 \\
24 & ---            & ---           && ---            & ---           && 15.32$\pm$0.18 & ---     & 0.19 \\
25 &  5.06$\pm$0.13 &  37.5$\pm$1.0 &&  5.37$\pm$0.14 &  39.8$\pm$1.0 && 10.43$\pm$0.19 &    1.06 & 0.14 \\
26 &  6.00$\pm$0.11 &  45.0$\pm$0.9 &&  8.95$\pm$0.11 &  67.2$\pm$0.9 && 14.95$\pm$0.16 &    1.49 & 0.11 \\
27 &  3.66$\pm$0.13 &  63.7$\pm$2.2 &&  4.42$\pm$0.13 &  76.8$\pm$2.2 &&  8.08$\pm$0.18 &    1.21 & 0.13 \\
\enddata
\footnotetext[1]{For epochs 16, 21, 22, and 24, the two sources are blended so only the total flux is given.}
\footnotetext[2]{When the source is not detected, the 3$\sigma$ upper limit is given.}
\end{deluxetable}

\section{Properties of the radio emission}

\subsection{Source flux and variability}

The radio source associated with V773 Tau A is found to be extremely variable from 
epoch to epoch (Table \ref{table_fluxes}) with a maximum recorded total flux (at epoch 
23) one hundred times larger than the minimum recorded total flux (at epoch 1). There
is a clear tendency (Figure \ref{fig_flux}) for the flux to be higher near
periastron, and lower around apastron, in agreement with the single-dish
results by Massi et al.\ (2002, 2006). The present data demonstrate for the
first time that this tendency holds for both sub-components (Aa and Ab) of the
system (Figure \ref{fig_flux}). The primary is found to be brighter than the
secondary in about 65\% of the cases, while the secondary in brighter than the
primary in the other 35\%. On average, the primary is about 10--15\% brighter
than the secondary. The brightness temperature of both components fluctuates
between 3 $\times$ 10$^{6}$ K and 2 $\times$ 10$^{8}$ K, indicating that the
emission is of non-thermal origin.

Short timescale (intra-epoch) variability is also clearly present in our
data. To characterize that variability, we have split each of our observations
into 1-hour chunks, and measured the flux of each sub-components of the system
in those short time intervals. A summary of the results is provided in Table
\ref{table_variability}, where we consider separately the situation near
periastron, and around apastron. The mean source flux ($\bar{f}_\nu$, measured
in 1 hour chunks) is typically 4 to 5 times higher near periastron than near
apastron, in agreement with our conclusions for epoch-integrated fluxes. The
relative variability of the source, as measured by the ratio of the
chunk-to-chunk r.m.s.\ to the mean is found to be between 40 and 120\% with 
no significant trend between periastron and apastron.

\subsection{Circular polarization}

Circular polarization was unambiguously detected at 6 epochs\footnote{It is
  important to recall that polarization calibration was only possible for the
  13 epochs corresponding to project BL146. Thus, our detection rate of
  circular polarization is roughly 50\%.} (Table \ref{table_pol}). The
polarization is associated with source Aa in two cases and with source Ab in
two other cases. The remaining two cases correspond to epochs 22 and 24 when
the sources are blended together. The levels of detected circular polarization
(between a few and about 10\%) are typical of gyrosynchrotron emission (Dulk
1985). The lack of circular polarization detections in many of our
observations presumably indicates that the topology of the magnetic field is
complex.

\subsection{Structure and size of the emitting region}

The radio emission detected in our 27 observations is largely confined to two
compact sources associated with the two stars in the system. There is only
marginal evidence for additional emission in the system --for instance in the
form of a faint bridge of emission between the stars at epoch 14 (see Figure
\ref{fig_epochs}). Deeper observations of the system should be obtained to
confirm the existence of such faint extended emission, and analyze its
properties in detail. It is clear, in any case, that such emission
statistically only represents a small fraction of the total emission.

A compact radio source associated either with component Aa or with component
Ab is detected a total of 43 times (23 times with source Aa and 20 times with
source Ab). Out of this total, the source is found to be completely unresolved
20 times (46.5\%), and resolved along only one direction in another 15 epochs
(35\%). The source is resolved along both directions (and, even then, only
marginally) in only 8 cases (18.5\%). The upper limit on the size of the
emitting regions along the unresolved dimensions is about 0.7 mas, whereas the
deconvolved linear size of the emission along the resolved dimensions is on
average 1 mas. Since the radii of both Aa and Ab are $R_*$ $\sim$ 2 \Rsun
(B2007), the radio emission detected here is almost entirely confined to
regions of radius 7$R_*$, and is often confined to regions of radius smaller
than 5$R_*$. These figures characterize the size scale of the magnetospheres
associated with Aa and Ab.

\medskip

In summary, the radio emission detected here is confined to very compact (5--7
$R_*$) regions. It is very variable, occasionally circularly polarized, and
has a very high brightness temperature ($T_b$ $\sim$ 10$^7$ K). All these
elements clearly indicate that the emission is primarily of gyrosynchrotron
origin (Dulk 1985).

\begin{deluxetable}{lrrrrr}
\tablecaption{Short-term variability.\label{table_variability}}
\tablehead{
\multicolumn{1}{c}{~~~~~~~~~~~~~~~~~Source~~~~~~~~~~~~~~~~~}&
\multicolumn{1}{c}{$\bar{f_\nu}$}&
\multicolumn{1}{c}{$rms$}&
\multicolumn{1}{c}{$f_\nu^{\rm ~max}$}&
\multicolumn{1}{c}{$f_\nu^{\rm ~min}$}\\%
[0.1cm]\cline{2-5}\\[-0.1cm]
\multicolumn{1}{c}{}&
\multicolumn{4}{c}{[mJy]}
}%
\startdata
\multicolumn{5}{l}{$0.3 \le \phi \le 0.7$ (Apastron)} \\
[-0.2cm]\\%
V773~Tau~A\dotfill  &  5.2 & 4.7 & 14.8 &   0.6 \\%
V773~Tau~Aa\dotfill &  2.7 & 2.1 &  5.8 &   0.3 \\%
V773~Tau~Ab\dotfill &  2.5 & 3.0 &  9.0 &$<$0.1 \\%
[-0.2cm]\\%
\multicolumn{5}{l}{$\phi \le 0.2$ or $\phi \ge 0.8$ (Periastron)} \\%
[-0.2cm]\\%
V773~Tau~A\dotfill  & 20.6 & 9.0 & 40.7 &   8.6 \\%
V773~Tau~Aa\dotfill &  8.4 & 3.3 & 12.0 &$<$0.2 \\%
V773~Tau~Ab\dotfill & 10.7 & 9.4 & 28.7 &   2.9 \\%
\enddata
\end{deluxetable}

\subsection{A conundrum}

We showed in Section 4.1 (see Figure \ref{fig_flux}, and Massi et al.\ 2002,
2006) that the radio flux is strongly dependent on the orbit phase of the
binary system, the typical flux near periastron being typically five times
higher than the typical flux near apastron. This clearly indicates that the
magnetic activity is enhanced when the stars approach one
another. Interestingly, however, the separation between the stars at
periastron is still about 2 mas or 30 $R_*$. Thus, even at periastron, the
magnetospheres of the two stars (of radii 5--7 $R_*$; Section 4.3) do not
overlap. How can the radio emission be so strongly dependent on the separation
between the stars if the radius of the magnetospheres is several times smaller
than the shortest separation between the stars? Even more puzzling, how can
the emission still remain almost completely concentrated to 5--7 $R_*$
magnetospheres near periastron, if the emission is so strongly enhanced by the
proximity between the stars?

A possible solution to this conundrum has recently been proposed by Massi et
al.\ (2008) who suggested the existence of two kinds of magnetic structures
around the stars in V773~Tau~A. The first kind corresponds to compact, closed
magnetic loops extending to only a few stellar radii. The second category are
semi-open structures similar to the so-called helmet streamers observed around
the Sun (e.g.\ Vourlidas 2006). Those streamers are anchored on the upper edge
of a compact closed loop at a few stellar radii above the stellar surface, but
they extend to about 30 stellar radii. Reconnections between these extended loops 
could occur when the stars of V773~Tau~A are near periastron, and naturally 
explain the associated radio flux increase. It is indeed noteworthy that the 
displacement between the stars and the radio sources observed near periastron 
(2--3 $R_*$; Section 3.1) is very similar to the height above the stellar surface at which
helmet streamers are expected to be anchored. We note, however, in our observations 
obtained near periastron, that even though the radio emission is slightly displaced from 
the stars, it is still clearly resolved into two compact structures each 5--7 $R_*$ across, 
but separated by more than 20 $R_*$. We find very little evidence for any emission 
between the stars. In particular,
we do not confirm the detection of the mirror magnetic points reported by Massi et al.\
(2008) in one of their epochs. 

\begin{deluxetable}{clr}
\tablecaption{Circular polarization.\label{table_pol}}
\tablehead{
\multicolumn{1}{c}{Epoch}&
\multicolumn{1}{c}{Source}&
\multicolumn{1}{c}{Polarization}\\%
 & & \multicolumn{1}{c}{[\%]}}
\startdata
15 &Ab  &10.5 $\pm$2.5 \\%
18 &Aa  &12.3 $\pm$3.8 \\%
22 &Aab & 3.5 $\pm$0.3 \\%
23 &Ab  & 1.3 $\pm$0.3 \\%
24 &Aab & 1.8 $\pm$0.5 \\%
26 &Aa  &11.1 $\pm$1.3 \\%
\enddata
\end{deluxetable}

\section{Conclusions}

In this paper, we have described and analyzed new multi-epoch VLBA
observations of the young stellar system V773~Tau~A. The data have been used
both to improve the determination of the physical orbit of the compact binary
at the center of the system, and to measure its trigonometric parallax. We
show that the distance to the system obtained from the a priori B2007 and
updated orbit model is fully consistent with the direct parallax
measurement. The proper motion of the barycenter of V773~Tau~A appears to be
accelerated, and the measured acceleration vector is in excellent agreement
with the expectation based on the inferred gravitational influence of
V773~Tau~B (B2011). 
Our results show that V773~Tau is at the same distance and shares the same 
kinematics as two other nearby young stars (Hubble~4 and HDE~283572) with 
VLBA parallax measurements. We argue that the mean distance to these three 
stars (131.4 pc) and their dispersion (2.4 pc) provide a good estimate of the distance 
and depth of the young population associated with this portion of Taurus (which
corresponds to the dark cloud Lynds~1495).

The radio emission detected here is confined to very compact (5--7 $R_*$)
regions associated with the two stars (Aa and Ab) in the system. It is very
variable, occasionally circularly polarized, and has a very high brightness
temperature ($T_b$ $\sim$ 10$^7$ K). All these elements clearly indicate that
the emission is primarily of gyrosynchrotron origin. Each time the system
passes near periastron, the radio emission increases by a factor of about 5,
and is slightly displaced from the stars.

\acknowledgements R.M.T. and W.H.T.V., acknowledges support by the Deutsche
Forschungsgemeinschaft (DFG) through the Emmy Noether Research grant VL
61/3-1. L.L. and L.F.R. acknowledge the financial support of DGAPA, UNAM and
CONACyT, M\'exico. L.L.\ acknowledges financial support from the Guggenheim 
Foundation and the von Humboldt Stiftung. The National Radio Astronomy 
Observatory is a facility of the National Science Foundation operated under 
cooperative agreement by Associated Universities, Inc.


\begin{thebibliography}{}

\bibitem[Boden et al.\ (2000)]{2000ApJ...536..880B} Boden, A.\ F.,
  Creech-Eakman, M.\ J., Queloz, D., 2000 \apj, 536, 880

\bibitem[Boden et al.\ (2007)]{2007ApJ...670.1214B} Boden, A.\ F.,
  Torres, G., Sargent, A.\ I., Akeson, R.\ L., Carpenter, J.\ M.,
  Boboltz, D.\ A., Massi, M., Ghez, A.\ M., Latham, D.\ W., Johnston,
  K. J., Menten, K.\ M., Ros, E., 2007, \apj, 670, 1214

\bibitem[Boden et al.\ (2011)]{}Boden et al.\ in prep., 2011

\bibitem[Briggs(1995)]{1995AAS...18711202B} Briggs, D.\ S., 1995,
  Bulletin of the American Astronomical Society, 27, 1444

\bibitem[Dzib et al.\ (2010)]{2010ApJ...718..610D} Dzib, S., Loinard,
  L., Mioduszewski, A.\ J., Boden, A.\ F., Rodr\'{\i}guez, L.\ F.,
  Torres, R.\ M., 2010, \apj, 718, 610

\bibitem[Dzib et al.(2011)]{2011ApJ...733...71D} Dzib, S., Loinard, L., 
Rodr{\'{\i}}guez, L.~F., Mioduszewski, A.~J., 
\& Torres, R.~M.\ 2011, \apj, 733, 71 

\bibitem[Duch\^ene et al.\ (2003)]{2003ApJ...592..288D} Duch\^ene, G.,
  Ghez, A.\ M., McCabe, C., Weinberger, A.\ J., 2003, \apj, 592, 288

\bibitem[Dulk (1985)]{1985ARA&A..23..169D} Dulk, G.\ A., 1985, \aap,
  23, 169

\bibitem[Feigelson et al.\ (1994)]{1994ApJ...432..373F} Feigelson, E.\
  D., Welty, A.\ D., Imhoff, C., Hall, J.\ C., Etzel, P.\ B.,
  Phillips, R.\ B., Lonsdale, C.\ J., 1994, \apj, 432, 373

\bibitem[Ghez et al.\ (1993)]{1993AJ...106..2005G} Ghez, A.\ M.,
  Neugebauer, G., Matthews, K., 1993, \aj, 106, 2005

\bibitem[Greisen (2003)]{2003ASSL..285..109G} Greisen, E.\ W., 2003, in
  Information Handling in Astronomy -- Historical Vistas, edited by A.\ Heck
  (Dordrecht: Kluwer Academic Publishers)

\bibitem[Hartmann et al.(1986)]{1986ApJ...309..275H} Hartmann, L., Hewett, 
R., Stahler, S., \& Mathieu, R.~D.\ 1986, \apj, 309, 275 

\bibitem[Hillenbrand \& White (2004)]{2004ApJ...604..741H}
  Hillenbrand, L.\ A., White, R.\ J., 2004, \apj, 604, 741

\bibitem[Kutner et al.\ (1986)]{1986AJ.....92..895K} Kutner, M.\ L.,
  Rydgren, A.\ E., \& Vrba, F.\ J., 1986, \aj, 92, 895

\bibitem[Leinert et al.\ (1993)]{1993A&A...278..129L} Leinert, Ch.,
  Zinnecker, H., Weitzel, N., Christou, J., Ridgway, S.\ T., Jameson,
  R., Haas, M., Lenzen, R., 1993, \aap, 278, 129

\bibitem[Lestrade et al.\ (1999)]{1999A&A...344.1014L} Lestrade,
  J.-F., Preston, R.\ A., Jones, D.\ L., Phillips, R.\ B., Rogers, A.\
  E.\ E., Titus, M.\ A., Rioja, M.\ J., Gabuzda, D.\ C., 1999, \aap,
  344, 1014

\bibitem[Loinard et al.\ (2005)]{2005ApJ...619L.179L} Loinard, L.,
  Mioduszewski, A.\ J., Rodr\'{\i}guez, L.\ F., Gonz\'alez, R.\ A.,
  Rodr\'{\i}guez, M.\ I., Torres, R.\ M., 2005, \apj, 619, L179

\bibitem[Loinard et al.\ (2007)]{2007ApJ...671..546L} Loinard, L.,
  Torres, R.\ M., Mioduszewski, A.\ J., Rodr\'{\i}guez, L.\ F.,
  Gonz\'alez-L\'opezlira, R.\ A., Lachaume, R., V\'azquez, V.,
  Gonz\'alez, E., 2007, \apj, 671, 546

\bibitem[Loinard et al.\ (2008)]{2008ApJ...675L..29L} Loinard, L.,
  Torres, R.\ M., Mioduszewski, A.\ J., Rodr\'{\i}guez, L.\ F., 2008,
  \apj, 675, L29

\bibitem[Martin et al.\ (1994)]{1994A&A...282..503M} Martin, E.\ L.,
  Rebolo, R., Magazzu, A., Pavlenko, Ya. V., 1994, \aap~282, 503

\bibitem[Mason et al.\ (2001)]{2001AJ....122.3466M} Mason, B.\ D.,
  Wycoff, G.\ L., Hartkopf, W.\ I., Douglass, G.\ G., Worley, C.\ E.,
  2001, \aj, 122, 3466

\bibitem[Massi et al.\ (2002)]{2002A&A...382..152M} Massi, M.,
  Menten, K., Neidh\"ofer, J., 2002, \aap, 382, 152

\bibitem[Massi et al.\ (2006)]{2006A&A...453..959M} Massi, M.,
  Forbrich, J., Menten, K.\ M., Torricelli-Ciamponi, G., Neidh\"ofer,
  J., Leurini, S., Bertoldi, F., 2006, \aap, 453, 959

\bibitem[Massi et al.\ (2008)]{2008A&A...480..489M} Massi, M., Ros,
  E., Menten, K.\ M., Kaufman Bernad\'o, M., Torricelli-Ciamponi, G.,
  Neidh\"ofer, J., Boden, A., Boboltz, D., Sargent, A., Torres, G.,
  2008, \aap, 480, 489

\bibitem[Mathieu et al.\ (2007)]{2007prpl.conf..411M} Mathieu, R.\ D.,
  Baraffe, I., Simon, M., Stassun, K.\ G., White, R., 2007, PRPL, 411

\bibitem[O'Neal et al.\ (1990)]{1990AJ....100.1610O} O'Neal, D.,
  Feigelson, E.\ D., Mathieu, R.\ D., Myers, P.\ C., 1990, \aj, 100,
  1610

\bibitem[Perryman et al.\ (1997)]{1997A&A...323L..49P} Perryman, M.\
  A.\ C., Lindegren, L., Kovalevsky, J., Hoeg, E., Bastian, U.,
  Bernacca, P.\ L., Cr\'ez\'e, M., Donati, F., Grenon, M., van
  Leeuwen, F., van der Marel, H., Mignard, F., Murray, C.\ A., Le
  Poole, R.\ S., Schrijver, H., Turon, C., Arenou, F., Froeschl\'e,
  M., Petersen, C.~S., 1997, \aap, 323, L49

\bibitem[Phillips et al.\ (1996)]{1996AJ....111..918P} Phillips, R.\
  B., Lonsdale, C.\ J., Feigelson, E.\ D., Deeney, B.\ D., 1996, \aj,
  111, 918

\bibitem[Rydgren et al.\ (1976)]{1976ApJS...30..307R} Rydgren, A.\ E.,
  Strom, S.\ E., Strom, K.\ M., 1976, \apjs 30, 307

\bibitem[Schmalzl et al.(2010)]{2010ApJ...725.1327S} Schmalzl, M., et al.\ 
2010, \apj, 725, 1327 


\bibitem[Torres et al.\ (2007)]{2007ApJ...671.1813T} Torres, R.\ M.,
  Loinard, L., Mioduszewski, A.\ J., Rodr\'{\i}guez, L.\ F., 2007,
  \apj, 671, 1813

\bibitem[Torres et al.\ (2009)]{2009ApJ...698..242T} Torres, R.\ M.,
  Loinard, L., Mioduszewski, A.\ J., Rodr\'{\i}guez, L.\ F., 2009,
  \apj, 698, 242

\bibitem[Torres (2010)]{2010arXiv1001.5144T} Torres, R.\ M., 2010, PhD
  Thesis, eprint arXiv:1001.5144

\bibitem[Vourlidas (2006)]{2006IAUS..233..197V} Vourlidas, A., 2006,
  Solar Activity and its Magnetic Origin, ed. V. Bothmer, \& A. Hady
  (Cambridge University Press), 197

\bibitem[Walter et al.(1988)]{1988AJ.....96..297W} Walter, F.~M., Brown, 
A., Mathieu, R.~D., Myers, P.~C., \& Vrba, F.~J.\ 1988, \aj, 96, 297 

\bibitem[Welty (1995)]{1995AJ....110..776W} Welty, A.\ D., 1995, \aj,
  110, 776

\bibitem[White \& Ghez (2001)]{2001ApJ...556..265W} White, R.\ J.,
  Ghez, A.\ M., 2001, \apj, 556,265

\bibitem[Woitas (2003)]{2003A&A...406..685W} Woitas, J., 2003, \aap,
  406, 685

\end{thebibliography}
\end{document}